\documentclass[aps,prl,twocolumn,amssymb,groupedaddress]{revtex4}

\usepackage{graphicx}

\begin{document}

\title{Atom Lithography using MRI-type feature placement}

\author{J.\ H.\ Thywissen}
\email[email: ]{joseph_thywissen@post.harvard.edu}
\author{M.\ Prentiss}
\email[email: ]{prentiss@fas.harvard.edu}
\affiliation{Department of Physics, Harvard University, Cambridge,
Massachusetts 02138, USA}

\date{\today}

\begin{abstract}
%
%
We demonstrate the use of frequency-encoded light masks in neutral
atom lithography.
We demonstrate that multiple features can be patterned across a
monotonic potential gradient. Features as narrow as 0.9\,$\mu$m are
fabricated on silicon substrates with a metastable argon beam.
Internal state manipulation with such a mask enables
continuously adjustable feature positions and feature densities 
not limited by the optical wavelength, unlike previous light masks.

\end{abstract}


\maketitle

%
%

%
Atom lithography has been used to make lines as small as 13\,nm
\cite{BellSmall,CrSmall}, to fabricate features with aspect ratios greater than
2:1 \cite{rhese,RIE}, and to pattern areas as large as 38\,cm$^2$
\cite{kjapl}. Atoms patterned with light masks are uniquely well-suited for
structured doping \cite{doping} and for fabrication demanding
long-range coherence \cite{nanoruler}.
However, feature spacings in light masks have been limited by the
length scale of the optical wavelength used
\cite{timp-prl,jabez-science,siuau-al,drodofsky,lison,SWQ,2D,engels,mutzel,brezger,jabezover8}.
In this Letter, we demonstrate a technique that extends light masks
beyond the feature density of optical lithography and to more complex
patterns than standing wave interferences.
%

%
Research in neutral atom lithography has been motivated by several
considerations. Neutral atomic beams with thermal kinetic energies
have short ($< .1$\,nm) de Broglie wavelengths, minimizing diffractive
resolution limitations. Neutral atoms are insensitive to stray
electric and magnetic fields, and long-range inter-particle
interactions are weak. A wealth of atom optics (mirrors, guides, etc.)
have been developed, including light masks, which, unlike physical
masks, do not sag and cannot be clogged or damaged.  Laser-accessible
internal structure of atoms allows efficient cooling, the storage of
internal energy, and complex manipulations.
State-sensitive atom lithography with metastable atoms has been
demonstrated with both positive- and negative-tone resists
capable of sub-10-nm resolution \cite{karl-sam,kjapl,passH}.

%
In this work we encode the desired position distribution 
using the {\it frequency}, instead of the intensity, of the light.
In the presence of an appropriate potential gradient, the optical resonant
frequency of the atom will be position-dependent.
Thus each spectral component of a laser beam interacts can interact
with the atoms at a specific but separate location.
The frequency encoding of spatial information
is commonly known from magnetic resonance imaging
(MRI), where protons (instead of atoms) are imaged (instead of
patterned) in a potential gradient by radio- (instead of optical-)
frequency radiation. 
We will refer to atom lithography with frequency-encoded light masks as
Atomic Resonance Lithography (ARL).
Thomas and coworkers have developed neutral atom measurement techniques
based on a more direct analogue to MRI \cite{thomas_th,thomas_exp}.
\begin{figure}[b!]
\includegraphics[width=8.5cm]{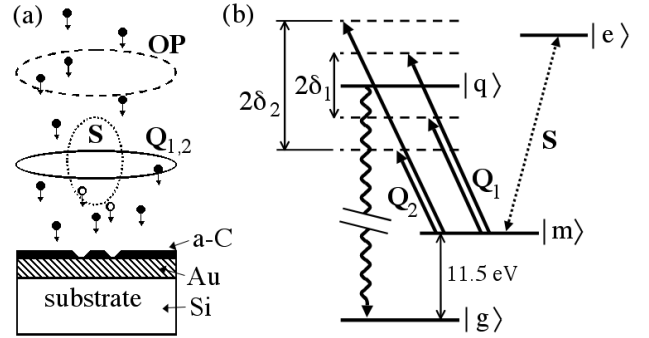}
\caption{The first of two light masks used for atomic resonance lithography.
{\bf (a)} Argon atoms strike a gold surface after passing through a
sequence of laser beams (not to scale) propagation parallel to the
surface: optical pump {\bf OP},
shift {\bf S}, and quench {\bf Q}. The light mask selectively quenches
some atoms to their ground state $|g\rangle$ (open circles), in which
state atoms are unable to activate the formation of resist on the
surface. At all other locations, metastable atoms $|m\rangle$ (filled
circles) activate the grown of carbonaceous resist (a-C) on the
substrate. {\bf (b)} Level diagram relevant to {\bf S} and {\bf Q}. An
on-resonant light shift beam ({\bf S}) is applied on a cycling
transition between the metastable state $| m \rangle$ and excited
state $|e\rangle$. Two pairs of detunings ({\bf Q}$_1$ and {\bf
Q}$_2$, detuned $\pm \delta_1$ and $\pm \delta_2$, respectively)
excite atoms to $|q\rangle$, from which atoms decay, in a radiative
cascade, to the inert ground state $|g\rangle$.
\label{fig:2f_lightmask} }
\end{figure}

Frequency encoding has several promising features.
Just as sub-100-$\mu$m features can be resolved with 100-meter
wavelength radio waves in MRI, far sub-optical (ie, sub-micron)
features and spacings can be created with optical resonances in atoms
\cite{thomas_th,thomas_exp,olshanii}.
As shown in Ref.~\cite{olshanii}, arbitrary patterns can be
frequency-encoded and generated. 
Furthermore, ARL can create two-dimensional patterns, just as MRI
can image two- \cite{lauterbur} or three-dimensional distributions.
Finally, frequency-encoded masks do not demand high transverse coherence of
the incident atoms. By contrast, although intricate holographic
patterns \cite{fujita,morinaga,shimizu} have been detected, the
required coherent sources lack the flux to pattern a surface:
experiments have used integrated surface densities on the order of
$10^9$ below a typical lithographic dose.
%

\begin{figure}[t!]
\includegraphics[width=8.5cm]{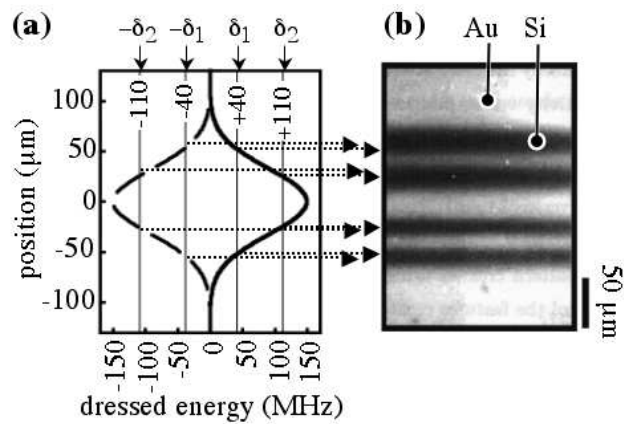}
\caption{{\bf (a)} Energy times $1/h$ versus position 
of the weak-field-seeking (solid line) and strong-field-seeking
(dashed line) dressed states of an atom in the {\bf S} beam. Thin solid
lines show the quench detunings $\pm \delta_1$ and $\pm
\delta_2$ {\bf (b)} Optical micrograph of a
gold-on-silicon substrate after exposure and etch. The dark lines are
exposed silicon, and correspond to locations where atoms were
quenched, thus preventing the formation of a resist and allowing gold
to be etched. These features are formed at resonant quench locations,
as indicated by dotted arrows between (a) and (b).
\label{fig:2f_lines}}
\end{figure}

%
Figure \ref{fig:2f_lightmask} describes 
a realization of ARL based on quenching directly within the 
potential gradient.
First, metastable argon atoms are optically pumped to the $|J=2, m_J=2
\rangle$ sublevel of the $4s[3/2]^o_2$ or $1s_5$ state. Not shown is
earlier optical quenching of the atoms in the $4s'[1/2]^o_0$ or 1s$_3$
state. If unquenched, these ``J=0'' atoms would contribute a
background rate of resist formation.
Next, the atoms pass through a zone of two
overlapping beams. A resonant $\sigma^+$ ``shift'' beam ({\bf S}) on
the 812\,nm $|J=2, m_J=2
\rangle$---$|J'=3, m_{J'}=3 \rangle$ cycling
transition overlaps with a $\sigma^-$ ``quench'' beam ({\bf Q})
containing four frequency components, $\pm\delta_1$ and
$\pm\delta_2$. As shown in Fig.~\ref{fig:2f_lines}a, these frequencies
are resonant at two stark shifts (i.e., four positions) in the
resonant shift beam. At these four positions, atoms are optically
pumped by {\bf Q} to the true atomic ground state $|g\rangle$. Note
that two frequencies (e.g., $\pm
\delta_1$) were required for complete optical pumping at 
each resonant position because atomic population existed at both the
strong- and weak-field-seeking dressed states.
This internal state manipulation creates a pattern on a surface by
exploiting the internal energy difference between the the ground state
and the $4s[3/2]^o_2$ metastable state of argon: atoms in $|m\rangle$
release 11.5\,eV of internal energy to activate the formation of a
resist on the surface (as described below), while atoms in $|g\rangle$
do not affect the surface.
Thus, the detunings $\delta_1$ and $\delta_2$ control where features
-- here, stripes of unprotected surface -- will appear.

%
Metastable argon (Ar*) atoms exposed the
substrate through the light mask with a flux density of $3 \times
10^{12}$\,cm$^{-2}$\,s$^{-1}$. Because the system is evacuated with
diffusion pumps, surfaces within the system are covered with several
monolayers of siloxane hydrocarbons.  When the metastable atoms hit
the surface, they transfer their internal energy to the physisorbed
hydrocarbons and induce a chemical change that results in the
formation of a durable material. The resist material remains on the
surface even after exposure to air and/or solution, and can thus be
used as a mask for etching. The beam apparatus and 
lithographic process used in this work have been described
in more detail elsewhere \cite{kjapl}.

A gold substrate was exposed for five hours through the optical mask
of Fig.\ \ref{fig:2f_lightmask}. The light mask parameters were as
follows. The {\bf S} beam had 5.9\,mW cylindrically focused to a
profile of 46\,$\mu$m by 925\,$\mu$m, creating a maximum stark shift
that was spectroscopically measured to be 150\,MHz.
Cylindrically symmetric 110-$\mu$m-wide quench beams {\bf Q}$_1$ and
{\bf Q}$_2$ each had 70\,$\mu$W of power, split between the pair of
detunings, $\pm \delta_1 = \pm 2 \pi \times 40$\,MHz and $\pm \delta_2
= \pm 2 \pi \times 110$\,MHz.
After exposure, the substrate was removed from the vacuum system and
etched for 7 minutes in a ferricyanide solution.

%
%
Figure~\ref{fig:2f_lines}b shows an optical micrograph of the etched
substrate.  At positions where either {\bf Q}$_1$ or {\bf Q}$_2$ was resonant,
metastable atoms were quenched, leaving the underlying substrate
unprotected against the subsequent gold etch. At all other positions, a
resist material was formed as described above, and gold remained on
the silicon substrate after etching. 
A pattern was formed across 2\,cm of the
substrate. Figure~\ref{fig:2f_lines}b was taken in the center of that
pattern, where atoms entered perpendicular to the light mask. The
comparison (indicated by dotted arrows) to Fig.~\ref{fig:2f_lines}a
shows that, as desired, there are four features: two on each side of
the Gaussian intensity profile of the {\bf S} beam.  We fit the sum of four
Lorentzians to the integrated reflectivity profile of the substrate to
determine more precise feature locations and positions. The lower
(upper) two lines have a half-width of 10$\pm$1\,$\mu$m
(14$\pm$1$\mu$m) and a center-to-center separation of 26$\pm$2$\mu$m
(34$\pm$2\,$\mu$m). The asymmetry between the upper and lower set of
lines was not expected, but could be explained by aberrations in the
cylindrical optics that formed the shift beam. The lower set of lines
are within experimental error of the expected linewidth of atomic
density, 9\,$\mu$m (as discussed below), and the expected feature
separation, 27\,$\mu$m. Note that the transfer function from atomic
flux density to reflectivity of the surface after etching is
nonlinear, such that purely atomic calculations may vary systematically
from the observed features.

%
The significance of the result shown in Fig.\ \ref{fig:2f_lines} is that
two lithographic features were created across each monotonic part of 
the potential. 
Unlike previous atom lithographic results with light masks, {\em the
wavelength of light does not determine the minimum feature
separation.}
%
The ultimate resolution of ARL can be limited by a variety of effects,
discussed Refs. \cite{thomas_th,olshanii} and summarized in the
following paragraphs.

The most fundamental limitation is the spectroscopic precision with
which each resonant transfer position is defined, $\delta x_{\rm sp} = {\hbar
\Gamma'}/{F}$, where $\Gamma'$ is the spectroscopic width of the
transition, and $F$ is the potential gradient.  In these experiments
we used laser light to create the potential gradient, but gradients
could also be created with magnetic \cite{thomas_exp} or electric
fields. For the realization of ARL discussed above, $\Gamma'$ is the
power-broadened and time-broadened quenching transition, with a
natural linewidth $\Gamma=2 \pi \times 5$\,MHz. Experimentally,
$\Gamma' = 2 \pi \times 20$\,MHz.  For the gradient $F/\hbar=2 \pi
\times 2.3$\,MHz/$\mu$m at the feature locations in
Fig.~\ref{fig:2f_lines}, we find $\delta x_{\rm sp}=9$\,$\mu$m.  Note
that for a Raman transition, as will be considered below,
$\Gamma'=v_L/w_0=1/\tau$, where $v_L=850$\,m/s is the longitudinal
velocity of the Ar* atoms, $w_0$ is the 1/e$^2$ intensity waist of
the Raman beams, and $\tau$ is the atom-light interaction time.

Atomic motion during the transfer can also limit the resolution. While
transferring an atom of mass $M$ to a state with a gradient $F$,
acceleration and diffraction limit the resolution to the order of ${{F
\tau^2}/{2 M}}$, where $\tau$ is again the transfer
time \cite{thomas_th}. However, the numerical factors depend on which
states are in a gradient and for how long.
In the case where $\Gamma'$ is determined by the interaction time of a
Raman transfer, and the pattern is formed by the atoms transferred to
the gradient, one can choose an optimal $\tau$ to give a resolution of
$\delta x_{\ell} = 2 \left( {\hbar^2}/{2 M F} \right)^{1/3}$.
For $^{40}$Ar and $F/\hbar=2 \pi \times 1$\, MHz/$\mu$m, the
optimal size is $\delta x_\ell = 100$\,nm. This limit is not, however,
fundamental: Olshanii {\em et al.} show that an appropriate choice of
probe frequency variation during atomic transfer (a ``magic phase'')
can correct for later evolution, such that the spectroscopic
resolution $\delta x_{\rm sp}$ is recovered \cite{olshanii}.

Free flight between the pattern formation and the substrate will allow
the atoms to diffract further. Given a free-flight time $T_{FF}$, the
diffraction limit is $\delta x_{\rm dif} = \sqrt{{\hbar T_{FF}}/{2 M}}$.
For instance, for $T_{FF} = 30 \Gamma^{-1}$, this limit is $\delta
x_{\rm dif} = 27$\,nm, smaller than the other limits given here.  Note
that this is a quantitative restatement of one of the desirable
qualities of atoms as a patterning constituent: their short de Broglie
wavelength $\lambda_{dB}$ minimizes the diffraction, 
$\sim \sqrt{\lambda_{dB} D}$, where $D$ is the mask-substrate
separation.

The collimation of the atomic beam can limit the resolution of
the pattern. Given an rms velocity spread $v_{rms}$, 
$\delta x_{\rm v} \approx v_{rms} (\tau+T_{FF})$.
For $\tau + T_{FF} = 30 \Gamma^{-1}$ and $v_{\rm rms} = 0.5$m/s, $\delta
x_{\rm v} = 0.4 \mu$m. Thus with a small (20\,nm) transverse coherence
length, sub-micron lithography can be performed.

\begin{figure}
\includegraphics[width=8.5cm]{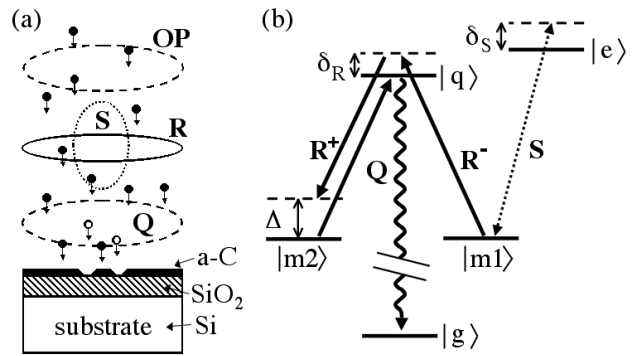}
\caption{ARL light mask using two-photon transfer and dark states. 
{\bf (a)} The geometrical arrangement of the light mask. The substrate
consists of a 15\,nm silicon-dioxide layer over a bulk silicon
layer. {\bf (b)} Atoms are prepared in the state
$|m1\rangle$ (optical pump transition not shown). At locations where
the stark shift of the {\bf S} beam matches the
differential detuning $\Delta$ of the Raman beams ({\bf R}$^+$ and {\bf R}$^-$),
atoms are transfered to $|m2\rangle$. In the subsequent quenching zone,
atoms in $|m2\rangle$ are optically pumped to $|g\rangle$ by {\bf Q}. 
\label{fig:arl_lightmask}}
\end{figure}

In order to improve our resolution, we performed a second experiment
with several experimental modifications. First, we improved the
quality of the optics forming the {\bf S} beam to reduce the waist by
a factor of 3. Next, as shown in Fig.~\ref{fig:arl_lightmask}, we used
a coherent transfer scheme by detuning the {\bf S} beam by 2\,GHz and
introducing Raman beams.  At resonant positions in the light mask, the
Raman beams transfer atoms from $|m1\rangle$ to $|m2\rangle$. In the
third interaction zone, only the atoms transferred to $|m2\rangle$ are
quenched \cite{statefn}.
With such a scheme, we realized a factor of
10 improvement in resolution, as described below. Furthermore, such a
coherent ARL scheme avoids momentum diffusion by spontaneous emission,
and can implement further resolution improvements such as the coherent
manipulations described in Ref.~\cite{olshanii}.
%

%
A silicon $\langle 100 \rangle$ substrate with a 15\,nm oxide layer
was exposed for six hours through the ARL mask depicted in
Fig.~\ref{fig:arl_lightmask}. The {\bf R} beam of 25\,$\mu$m by
103\,$\mu$m had $\sigma^-$ and $\sigma^+$ components with powers of
65\,$\mu$W and 60\,$\mu$W and detunings of $+61.1$\,MHz and
$+104.2$\,MHz, respectively. A 1.3\,G quantization field reduced the
differential detuning by 5.4\,MHz, such that the net Raman detuning
was $\Delta/2\pi=-37.7$\,MHz. The shift beam {\bf S} was detuned by
$\delta_S/2\pi = 2.00$\,GHz above resonance, had $6.9\pm0.3$\,mW of
power, and a profile of 15$\pm1$\,$\mu$m by 250\,$\mu$m. The
$\sigma^+$ polarized {\bf S} beam interacted with both $|m1\rangle$ and
$|m2\rangle$, but gave a differential gradient $F/\hbar=2 \pi \times
6.4$\,MHz/$\mu$m due to the difference in Clebsh-Gordan coefficients:
1 and $\sqrt{2/5}$, respectively (note that
Fig.~\ref{fig:arl_lightmask} shows only the transition $|m1\rangle-|e\rangle$,
for simplicity).
The third interaction zone had a 350\,$\mu$W quench beam ({\bf Q}) with a
waist of 210\,$\mu$m, such that about 12 photons are scattered from
atoms in $|m2\rangle$. Because $|m1\rangle$ was in the $|2,+2\rangle$,
it was dark with respect to the $\sigma^+$ polarized quench.
Untransferred atoms arrived at the substrate in the metastable state
$|m1\rangle$, and could therefore activate resist formation on the substrate.
\begin{figure}
\includegraphics[width=8.5cm]{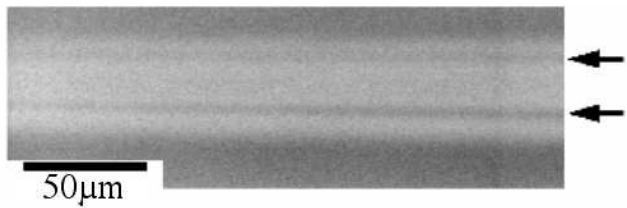}
\caption{SEM image of a Si $\langle100\rangle$ substrate patterned by
the light mask shown in Fig.~\ref{fig:arl_lightmask}. The darker lines
(indicated by arrows) are places at which metastable atoms were
quenched, thus preventing the formation of a resist material and
allowing the silicon to be etched. \label{fig:arl_lines}}
\end{figure}

%
After exposure, the pattern was transferred to
the substrate using both dry and wet etches. First, the sample was
etched for 35\,s in a CHF$_3$ reactive ion etch with parameters
42\,mTorr chamber pressure, 100\,W forward power, 10\,W reflected
power, -842\,V self-bias, and 25.0\,sccm flow rate. Second, the sample
was etched for 50\,s in 2\% HF to prepare an unoxidized silicon
face. Finally, the sample was etched for 14\,min in 20\% KOH at
22\,C \cite{RIE}.

Figure~\ref{fig:arl_lines} shows a scanning electron micrograph of the
etched sample. The features created by ARL are indicated by
arrows. Two features are created with a single Raman detuning
because the Stark shift is equal to $\Delta$ at a
location to either side of its maximum. The dark lines against a
bright background indicate places at which silicon was preferentially
etched. The sample was patterned across several millimeters of
length. At center, for the run parameters above (gradient $F/\hbar$=$2 \pi
\times 6.4\pm0.5$\,MHz$/\mu$m and $\tau=29\pm2$\,ns),
the expected feature separation is $19.0\pm1.5$\,$\mu$m, and the
spectroscopic limit is $\delta x_{\rm sp} = 0.86\pm0.07$\,$\mu$m.
This agrees well with our measurements: a minimum spacing
of $(20.0 \pm 0.4)\,\mu$m and a rms width of $(0.9\pm0.1)\,\mu$m.
%

%
%
These fabricated features are thirty times smaller than those that
have been detected in holographic masking experiments
\cite{morinaga}, but still two orders of magnitude larger than the
smallest features fabricated with atom lithography
\cite{BellSmall,CrSmall,rhese,RIE}.
Were the light gradient to be increased, feature sizes in the second
ARL scheme demonstrated would be limited by the velocity spread of the
atomic beam, at $\delta x_{\rm v} = 0.4$\,$\mu$m.  Therefore, the
spectroscopically limited feature size observed in
Fig.~\ref{fig:arl_lines} is a factor of two above the minimum
observable linewidth for the collimation of our atomic beam.  By
comparison, the resolution limit of an optimized light mask with a
gradient of $F/\hbar = 2 \pi \times 6.4$\,MHz/$\mu$m is $\delta x_\ell
= 54$\,nm for Ar*.

In conclusion, we have demonstrated the use of frequency encoding in
optical masks for atom lithography.
Using a two-photon internal state transfer, we create features in
silicon with an rms width of $0.9\pm0.1$\,$\mu$m.
With a mask based on optical pumping, we show that feature position
can be chosen continuously throughout monotonic potential
gradients. Such a method could be used to form intricate, sub-micron,
two-dimensional patterns with currently available sources.
Lithography using internal manipulation schemes could also be extended
to alkali atoms by state-selective ionization and
deflection. Unionized alkali atoms, like unquenched metastable atoms,
would be left to pattern an atom resist (e.g., \cite{CsSAMs}).

\begin{acknowledgments}
The authors thank N.\ Dekker and K.\ S.\ Johnson for contributions to
the early stages of this project. We also thank F.\ Altarelli, G.\
Zabow, and S.\ Coutreau for assistance. This work was supported in
part by the NSF Grant No.\ PHY-9876929, MRSEC Grant No.\
DMR-9809363, and the Hertz Foundation.

\end{acknowledgments}
%
%
%

%
%
\end{document}